# Impedance spectroscopy studies on lead free $Ba_{1-x}Mg_x(Ti_{0.9}Zr_{0.1})O_3$ ceramics

S. Ben Moumen[1], A. Neqali[1], B. Asbani[2], D. Mezzane[1*], M. Amjoud[1], E.Choukri[1], Y. Gagou[3], M. El Marssi[3], I. Lukyanchuk[3]

[1] LMCN, F.S.T.G Cadi Ayyad University Marrakech, Morocco

[2] Institut des Matériaux Jean Rouxel (IMN), CNRS UMR 6502 – Université de Nantes, France

[3] LPMC, Université de Picardie Amiens, France

*Corresponding author, E-mail address: daoudmezzane@gmail.com (D. Mezzane),

Faculty of sciences and technology, Av. Abdelkarim Elkhattabi B.P 549, Department of applied physics, Marrakech, Morocco.
Tel: +212 682860998

**Abstract**

$Ba_{1-x}Mg_x(Ti_{0.9}Zr_{0.1})O_3$ (x= 0.01 and 0.02) ceramics were prepared using the conventional solid state reaction. Rietveld refinement performed on X-ray diffraction patterns indicates that the samples are tetragonal crystal structure with *P4mm* space group. By increasing Mg content from 1 to 2% the unit cell volume decreased. Likewise, the grains size is greatly reduced from 10 μm to 4 μm. The temperature dependence of dielectric constants at different frequencies exhibited typical relaxor ferroelectric characteristic, with sensitive dependence in frequency and temperature for ac conductivity. The obtained activation energy values were correlated to the proposed conduction mechanisms.

## 1. Introduction

Ferroelectric materials with perovskite $ABO_3$ structures have received much attention due to their excellent functional properties [1,2]. These materials are widely used in electrical devices such as actuators, accelerators, ultrasonic generators, transducers and sensors [3,4]. Over the last few decades, enormous efforts have been devoted to the development of eco-friendly ceramics to avoid environmental pollution resulting from volatilization of lead monoxide PbO [5-9]. To overcome this drawback, researchers have developed new materials environmentally friendly. Among them, the most recognized lead-free candidates, is $Ba(Ti_{1-x}Zr_x)O_3$ ceramics family, a solid solution of $BaTiO_3$ and $BaZrO_3$ perovskite compounds. This family has been studied widely due to its high dielectric constant [10]. Physical properties of these materials depend strongly on the amount of Zr substitution [11]: when $0 < x < 0.2$ these materials have a Curie temperature above room temperature and they exhibit a normal ferroelectric behavior and good dielectric proprieties. When $x > 0.3$, the Curie temperature is below room temperature and a relaxor behavior is manifested [12]. Based on the attractive properties exhibited by $Ba(Zr_{0.1}Ti_{0.9})O_3$ (BZT) ceramics as their high dielectric constant, low dielectric loss, wide dielectric curve and a Curie temperature above room temperature [13], many research groups have studied the modified $Ba_{1-x}Y_x(Zr_{0.1}Ti_{0.9})O_3$ where Y is a bivalent metallic ion [14], owing to their promising applications with introducing bivalent metals on the A site of the perovskite structure. They have been interested in the study of electrocaloric effect, piezoelectricity by introducing metalic and alkaline ions metals such us Mg, Sn, Ca, La and Sr [15-19]. But rare are those who studied the dielectric and impedance spectroscopy properties of these materials. Reports in the literature show that the magnesium isovalent ion-substitution on $BaTiO_3$ with $Mg^{2+}$ ion in the A site lead to an enhancement of the dielectric properties [20]. On the other hand, few others show the properties of $Mg^{2+}$ ion substituted at the B site of $BaTiO_3$ ceramics [21, 22]. Furthermore it is well known in the literature that $Mg^{2+}$ doping or MgO addition could be used to reduce the dielectric loss in ceramics [23, 24]. Magnesium oxide MgO is also known to be an effective grain size inhibitor in many functional and structural ceramics. The aim of the present paper is to study the electrical properties of small amount Mg-doped BZT perovskite ceramics $Ba_{1-x}Mg_xZr_{0.1}Ti_{0.9}O_3$ (BMZT).

The impedance and dielectric data of BMZT (x=0.01 and x=0.02) are analyzed to understand the influence of small amount of Mg on the electrical relaxation of these ceramics. Likewise, the results of ac- and dc- conductivity measurements on these materials are studied.

## 2. Experimental procedure

Two polycrystalline samples of BMZT family with small amount of Magnesium x=0.01 and x=0.02, referenced BMZT1 and BMZT2 respectively, were synthesized using the conventional solid state reaction method. Stoichiometric amounts of high purity $BaCO_3$, (VWR, 99.99%), MgO, (Aldrich, 99.99%), $TiO_2$, (VWR, 99.5%), and $ZrO_2$, (Alfa Aesar 99%), were used as the starting materials. The precursors were mixed and grounded in ethanol medium using an agate mortar till obtaining a homogeneous mixture. The obtained slurry was dried and then calcined at 1250 °C for 16 hours. The calcined powders were mixed again and compacted into disks without using any binder. The pellets were sintered at 1450°C for 5 hours. The crystal structure was determined by an X-ray diffractometer (X'Pert PRO) using Cu-Kα (wavelength λ=1.5405 Å) radiation operated at 40 kV and 40 mA, the scans were recorded in a range from 10° to 80° with the step size of 0.017°. The cell parameters were refined from the diffraction patterns using the Fullprof software [25]. The morphologies of grain boundaries and surfaces were observed by scanning electron microscope (Tescan Vega 3 SEM). Dielectric properties were performed using an HP 4284A precision impedance analyzer.

## 3. Results and discussion

*3.1 Structural study*

X-ray powder diffraction patterns were refined and the obtained results on both samples are represented in Fig. 1. The diffraction peaks are sharp bearing out the good crystallinity of the BMZT powders. The structural analysis showed pure and single perovskite phases without any evidence of a detectable crystalline impurity. In order to confirm the crystallographic phases and the incorporation of $Mg^{2+}$ ion in the A-site the structural refinement was performed assuming a tetragonal perovskite structure with *P4mm* (N° 99) space group where (barium/magnesium) and (titanium/zirconium) ions occupy the 1a(0,0,z) and 1b(1/2,1/2,z) special positions respectively, and the oxygen anions occupy the 2c(0,1/2,z) and 1b positions. All the reflexion lines were successfully indexed. The occupancy rates of $Mg^{2+}$ in the barium site and $Zr^{4+}$ in titanium site were refined. The obtained values are in good agreement with the experimental data. A decrease in the cell volume is observed with the increase of the Mg amount, due to the substitution of smaller radius ion $Mg^{2+}$ (1.72 Å) into $Ba^{2+}$ (2.78 Å) sites [26]. The small differences between calculated and experimental patterns are plotted in Fig. 1 and the obtained reliability factor Rp, Rwp, atom positions, Biso and values of the fit goodness indicator $\chi^2$ are gathered in Table. 1. All these parameters demonstrate a satisfactory refinement.

**Fig.1.**

**Table.1.**

*3.2 SEM analysis*

The microstructure of BMZT modified ceramics with small Mg concentrations, x=0.01 and x=0.02, are shown in Fig. 2. The micrographs reveal relatively dense and uniform microstructure. They show the decrease of grain size when Mg mole fraction increases. The same behavior was observed previously in $(Ba_{1-x} Mg_x)(Ti_{0.98}Zr_{0.02})O_3$ [26]. The average grain sizes deduced from the histograms (see the insets) are found to be around 10 μm and 4 μm for BMZT1 and BMZT2, respectively.

**Fig.2.**

*3.3 Dielectric measurements*

Temperature dependence of dielectric permittivity ($\varepsilon_r$) and loss factor ($tan (\delta)$) measured at various frequencies for BMZT1 and BMZT2 ceramics are presented in Fig. 3. Interestingly, we notice that better dielectric properties are obtained with increasing of Mg content in BMZT. Indeed, important room temperature dielectric constant ($\varepsilon_r$ = 4200) at frequency 100 kHz with low dielectric loss value of *tan (δ)* was acheaved in BMZT1 comparing with the value ($\varepsilon_r$ = 3000) found in BZT by S. Ye et al [27] at the same frequency. With increasing frequency the values of dielectric constant decrease and the dielectric peak temperature ($T_m$) shifts toward high temperatures. It was found to be 52°C for BMZT1 and 37°C for BMZT2, at 10 kHz. For the parent undoped Ba(Ti$_{0.9}$Zr$_{0.1}$)O$_3$ compound, the dielectric constant maximum was observed at T$_m$=70 °C [27] at the same frequency value. That evidenced the decrease of T$_m$ with the increase of Mg content in BZT matrix, which could be due to the amphoteric character of Mg doping [28]. We observed low dielectric loss (*tan (δ)*) in the ferroelectric phase of both samples, which increases sharply in the paraelectric phase. This is more significant at high temperatures and at low frequencies, the signature of electrical conductivity.

**Fig.3.**

Furthermore, in the temperature range 150-400°C the dielectric permittivity does not follow the classical Curie-Weiss law.

$$\varepsilon' = \frac{C}{(T-T_0)} \qquad (1)$$

Moreover, the permittivity increases sharply with temperature in this range. The dielectric loss *tan (δ)* of both samples was relatively low near the ferroelectric-paraelectric transition, independently of the frequency. However, it showed a maximum near the phase transition temperature followed by increase with temperature. These results can be considered as a dielectric anomaly which is simply due to the different electrical properties of the bulk and the surface boundary layer [29], combining the conduction and polarization process. This anomaly can be attributed to the dielectric relaxation. Similar results have been reported in BaTiO$_3$ [30]. In order to study the nature of the phase transition, a detailed analysis was done based on Santos-Eiras equation [31].

$$\varepsilon'_r(T) = \frac{\varepsilon'_m}{1+\left(\frac{T-T_m}{\Delta}\right)^\gamma}, \qquad (2)$$

where $\varepsilon_m$ is the maximum of dielectric permittivity, $\Delta$ the parameter that defines the degree of diffuseness of the transition, γ indicates the character of the phase transition: if $1 < \gamma < 2$ the transition is "incomplete" this means that the interaction between nucleating ferroelectric clusters in paraelectric matrix begins far above $T_m$ and plays an important role in its diffusive character of transition. The results of the adjustments using the Santos-Eiras equation are presented in Fig. 4. The values of the fitting parameters $\varepsilon_m$, $T_m$, $\Delta$, and $\gamma$ are summarized in Table. 2 for all the studied compositions. The fit shows a good agreement between the theoretical and experimental curves which validate the applied model.

**Fig.4.**

**Table.2.**

The results of model adjustment show the decrease of $\varepsilon_m$ when frequency increases, whereas $T_m$ slightly increases exhibiting the relaxor behavior in both studied ceramics. The obtained values of $\Delta$ show an increase of the diffuse character of the transition with increase in Mg amount, maybe due to the cationic disorder in the system. $\gamma$ is equal to 1.49 for the parent compound elaborated with the same technique [11], [32]. Whereas the values of $\gamma$ are found to be higher than 1.6 for BMZT1 and BMZT2 indicating that the transitions are of incomplete diffuse type. The increase of $\gamma$ value with the increase of Mg amount indicates the increase of diffusivity in BZT matrix.

*3.4. Impedance spectroscopy*

The impedance data of the investigated samples are presented in the Nyquist diagram (Z" vs. Z') for several representative temperatures in the frequency range 20 Hz- 1 MHz. Complex impedance data were fitted by using Z-view and plotted in Fig. 5. It is observed that with the increase in temperature the slope of the lines decreases and the curve moves towards real (Z') axis indicating the increase in conductivity in the samples. The first semicircle, at low frequency, represents the grain boundary contribution. The second one positioned at high frequency region corresponds to the grain or bulk properties. Moreover, the electrical behavior of the samples is well represented by two resistance-constant phase element (R//CPE) equivalent circuits connected in series, as shown in the inset of Fig. 5(b).

**Fig.5.**

The resistance and capacitance of BMZT1 and BMZT2 for the bulk and grain boundary were deduced from the fitting of Nyquist plot. The relaxation time ($\tau$) was then calculated using the equation.

$$\tau = (R*Q)^{1/n} \tag{3}$$

where R is the resistance, Q is the capacitance of the CPE element and n is the empirical exponent. Bulk characteristic value was specified by "g" and grain boundary by "gb". The values of resistances and capacities for both grain and grain boundaries are gathered in Table 3.

**Table.3.**

The relaxation time follows Arrhenius equation:

$$\tau = \tau_0 \exp(E_a/K_B T) \tag{4}$$

where $\tau_0$ is the pre-exponential factor, Ea is the activation energy and $K_B$ is the Boltzmann constant [33]. The activation energies are calculated from the slopes of the fitted straight lines taken from the plot of $\tau$ versus 1000/T presented in Fig. 6. The calculated activation energies for the grain are found to be 1.03 eV and 1.17 eV and for the grain boundary are 1.02 eV and 1.15 eV for BMZT1 and BMZT2 respectively.

**Fig.6.**

In Fig 7 (a) and (b) we present the optimal fitted curves that show good accordance with experimental variation of real part of impedance (Z′) as a function of frequency at different temperatures for BMZT1 and BMZT2 samples, respectively. A monotonic decrease of Z′ value with temperature can be observed. Z′ value is higher at lower temperature in the low frequency region and it decreases gradually with the increase in frequency which indicates the increase in ac conduction ($\sigma_{ac}$) in the samples. The value of Z′ increases with increasing Mg concentration from 1% to 2 %. Its value merges out in the high frequency region regardless of temperature, This could be due to the release of space charge [34, 35]. as a result of reduction in the potential barrier properties in the material at the presence of Mg.

The variation of the imaginary part of impedance (Z″) as a function of frequency at different temperatures for BMZT1 and BMZT2 samples are presented in Fig 7 (c) and (d), respectively. The $Z''_{max}$ peaks shifts and Typical broadening was observed at higher frequencies with increasing temperature and then, merges in the higher frequency region irrespective of the temperature. Moreover, the value of $Z''_{max}$ shifts to higher frequencies with increasing temperature and it merges in the higher frequency region irrespective of the temperature. This may be an indication of the temperature dependent relaxation. The broadening behavior shows the presence of relaxation process, which is temperature dependent [36]. The relaxation process may be also due to the presence of immobile species - electrons at low temperature and defects - vacancies at higher temperatures. In BMZT1 and BMZT2 samples, the merging of Z″ max peaks occurs below 1 MHz. Two peaks can be observed, the first polarization phenomenon at low frequency is correlated with grain boundary, while the second one is correlated with the bulk or grain contribution. These results suggest that there are two major distributions with different relaxation times.

**Fig.7.**

In general, the origin of ac conductivity $\sigma_{ac}$ is mainly caused by the hopping of ions. The ac conductivity gives an overview of the different relaxation mechanisms based on its frequency and / or temperature dependent characteristics. Fig. 8 shows the variation of the ac conductivity ($\sigma_{ac}$) with the frequency, starting out from its extrapolation at the zero frequency ($\sigma_{dc}$), and ultimately attaining a high frequency almost a plateau from 10 kHz to 1 MHz. The conductivity values are between $10^{-2}$ (S/m) and $10^{-4}$ (S/m) in the frequency range 20 - $10^6$ Hz. For both BMZT1 and BMZT2. A dispersion nature of $\sigma_{ac}$ is observed, it shifts slightly to high frequency with increasing temperature. This may be due to the polarization of the electrode. In the conductivity spectrum, we discern three parts: the high frequency plateau regime (region I), middle frequency dispersive regime (region II) and the low frequency (dc) plateau regime (region III). These characteristics can be explained using the jump relaxation model [37]. The plateau in (region III) at low frequency for high temperature is attributed to the long-range translation motion of ions where the field cannot perturb the hopping mechanism of the charged particles. This part of the spectra can be described by the Jonscher's power law [38] according to the equation:

$$\sigma_{ac}= \sigma_{dc}+A\omega^n \qquad (5)$$

where n is the frequency exponent evolving in the range of $0 < n < 1$. A and n are thermally activated quantities, hence electrical conduction is a thermally activated process. According to Jonscher [36], the origin of the frequency dependence of conductivity lies in the relaxation phenomena arising from mobile charge carriers. When mobile charge carriers hop from one localized site into a neighboring vacant one, it remains in a state of displacement between two potential energy minima, which includes contributions from other mobile defects after long time period. The defect could relax until the two minima in lattice potential energy coincide with the lattice site. The term $A\omega^n$ contains the dependence on ac current and characterizes any dispersion phenomenon. Funke [38] explained that the value of n can have a physical meaning, if $n \leq 1$, then the charge carriers take a translation motion with sudden hopping, when $n > 1$, the charge carriers make small local hopping without leaving their neighboring sites [35]. Furthermore, at high frequencies, two competing relaxation processes take place (i) unsuccessful hopping in which the hopping ion returns to its initial position in this case the process is called the forward–backward hopping process and (ii) the successful hopping where the neighboring ions go relaxed and the hopping ion stays on the new site. In the conductivity response, each individual hop is observed when the activity time is short enough. This is sensitive at high and constant hopping conductivity in (region I). In the dispersive regime (region II) only the hops which are successful within a time interval of duration equal to $1/2\pi f$ will contribute to conductivity. The ratio of successful to unsuccessful hopping results in more dispersive conductivity. The increase of conductivity with frequency in region II is thus qualitatively explained [38].

**Fig.8.**

Figure 9 depicts the behavior of the ac conductivity in the ceramics versus temperature at three chosen frequencies, a strong temperature dependence is observed in both cases, we notice that the plots can be fragmented into three regions, a ferroelectric one designed by FEI and two paraelectric regions PEI and PEII characterized with different slopes revealing different conduction mechanisms with different values of activation energies (Ea). The noticeable change in the slopes of BMZT1 and BMZT2 observed between the FEI and PEI regions is probably due to a difference on the readjustment of the lattice during the ferroelectric paraelectric crystalline phase transition. This confirms the diffuse character of the transition observed earlier on Fig. 4. with increasing Mg amount. A straight line is observed at high temperatures PEII revealing a sharp increase in conductivity, ac conductivity converges at high temperature independently of frequency indicating saturated regime of ac conductivity. The activation energies were calculated using the Arrhenius equation:

$$\sigma_{ac} = \sigma_0 \exp(-E_a/K_B T), \tag{6}$$

where $\sigma_0$ is the pre-exponential factor, Ea is the activation energy and $K_B$ is the Boltzmann constant. Some representative activation energies at various frequencies are gathered in Table 4 (a) and (b).

**Fig.9.**
**Table.4.**

## 4. Conclusions

Lead free $Ba_{1-x}Mg_xTi_{0.9}Zr_{0.1}O_3$ ceramics with different (x=0.01, 0.02) were prepared from powders synthesized using the conventional solid state reaction. Several properties of the materials have been studied using different techniques. By X-rays diffraction technique we confirmed tetragonal perovskite structure symmetry described in *P4mm* space group. It was also observed that the increase of the Mg content caused the decrease in lattice parameters, unit cell volume and the grain size sharply decreased from ~10 μm to ~4 μm, leading to a shift down of Curie temperature. Impedance spectroscopy and electrical conductivity analysis were performed. A diffuse dielectric anomaly was observed in the temperature range 150-400 °C attributed to the dielectric relaxation process. Both ceramics have mainly two contributions of grain and grain boundary, the relaxation activation energies were calculated. The obtained results showed conduction process occurs in both grains and grain boundaries. Three conductivity regions were evidenced via the ac conductivity analysis. Global conductivity is supported by transport of charge carriers over a long distance in the low frequency region followed by a slop change from the grain boundary capacitance contribution, and the second plateau is attributed to bulk conductivity.

## Acknowledgements

This work was supported by CNRST Priority Program PPR 15/2015 and H2020-MSCA-RISE-2017-ENGIMA action.

**Figures**

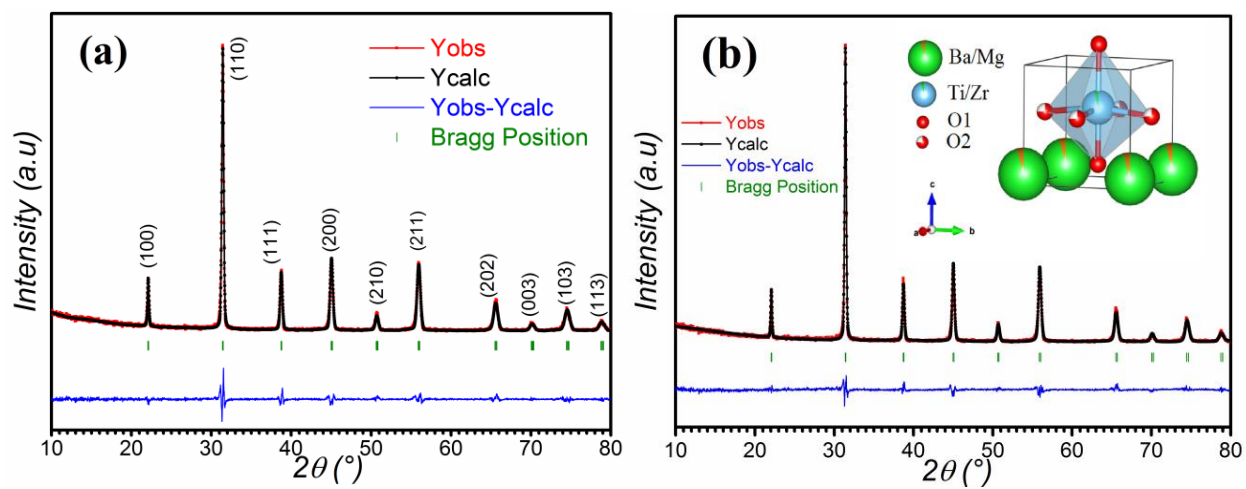

Fig.1. (Color online) Refined X-ray powder diffraction patterns of (a) BMZT1 and (b) BMZT2 samples. The inset shows the refined crystal structure.

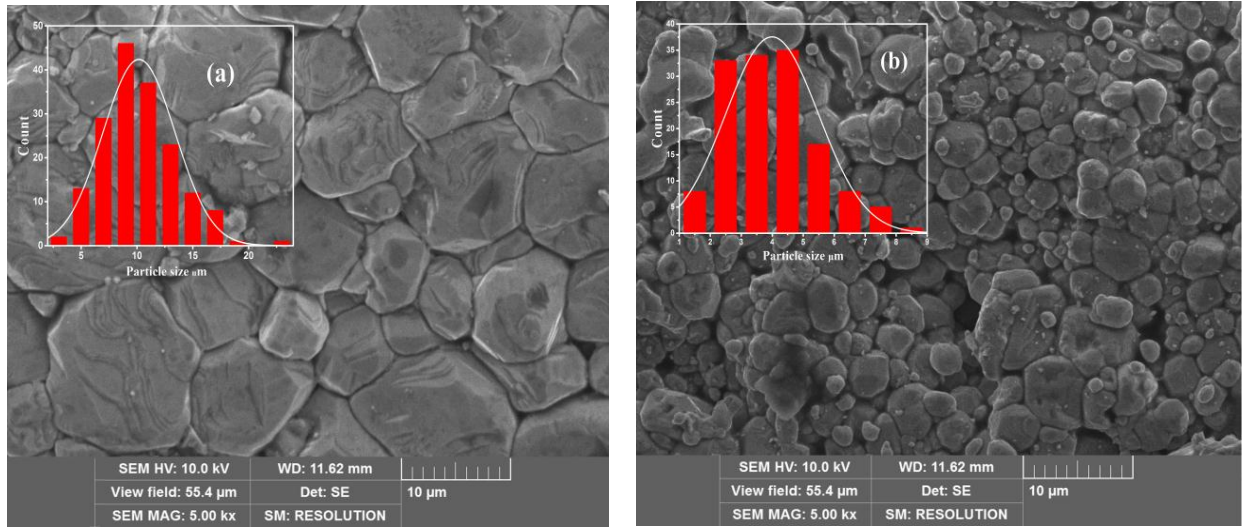

Fig.2. SEM micrographs of (a) BMZT1 and (b) BMZT2 samples, the insets show the grain size distribution histograms adjusted to a normal law.

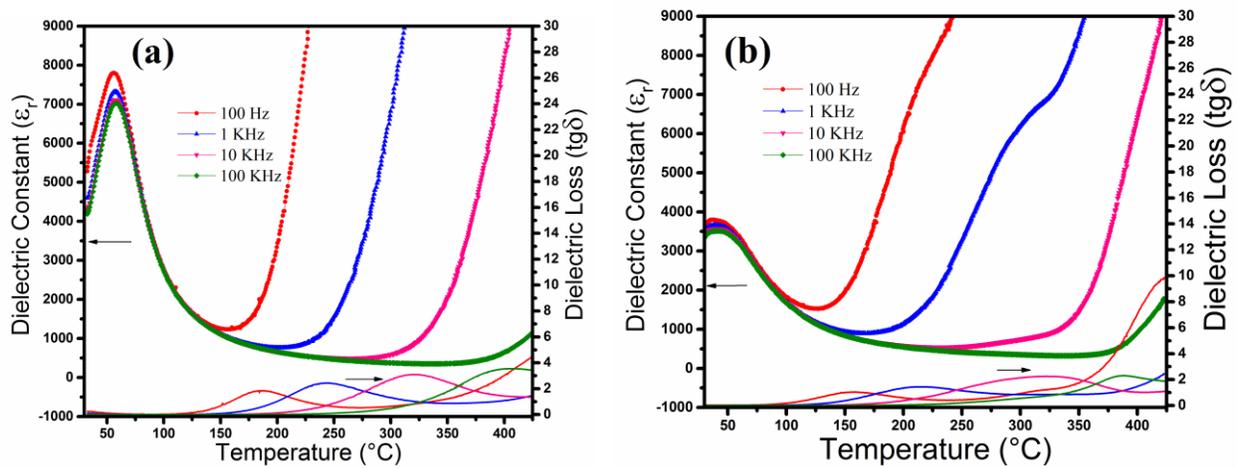

Fig.3. (Color online) Temperature dependence of dielectric constant and loss factor of (a) BMZT1 and (b) BMZT2.

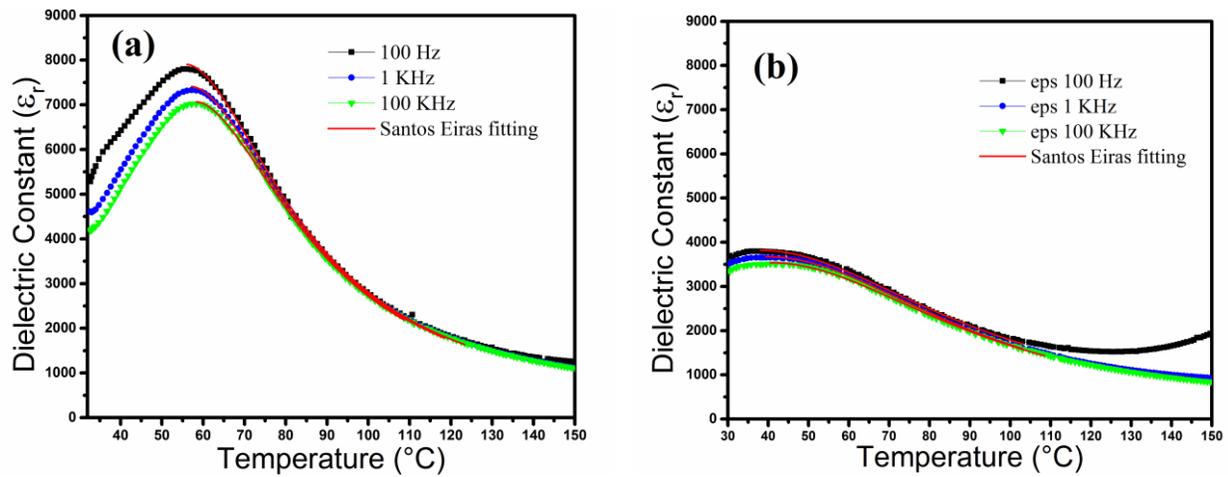

Fig.4. (Color online) Temperature dependence of dielectric constant of (a) BMZT1 and (b) BMZT2, the solid lines present the theoretical Santos-Eiras fit.

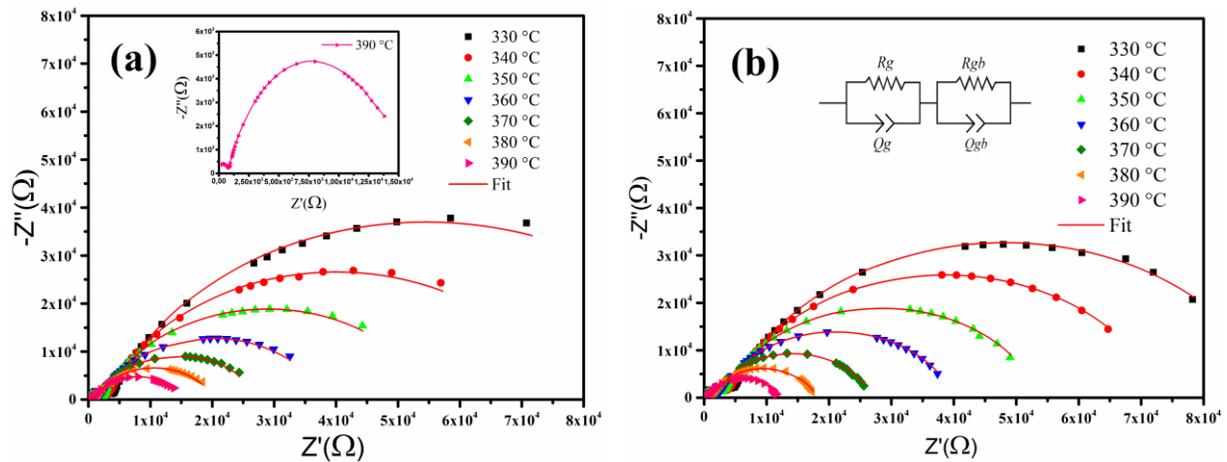

Fig.5. (Color online) Nyquist diagram of (a) BMZT1 and (b) BMZT2 at different temperatures. Insets show (a) zoom of Z''-Z' plot at 390°C and (b) the appropriate equivalent electrical circuit.

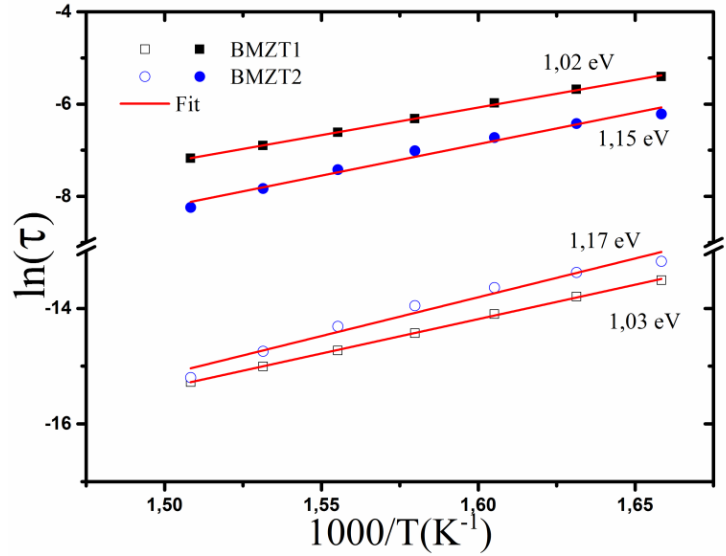

Fig.6. (Color online) Variation of relaxation time with inverse of temperature for grains (open markers) and grain boundaries (solid markers) for BMZT1 and BMZT2.

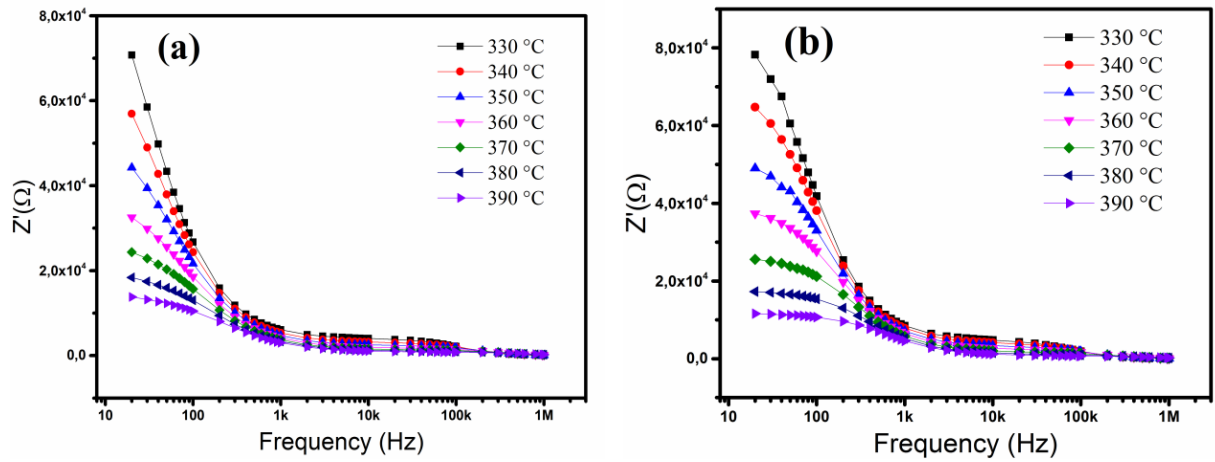

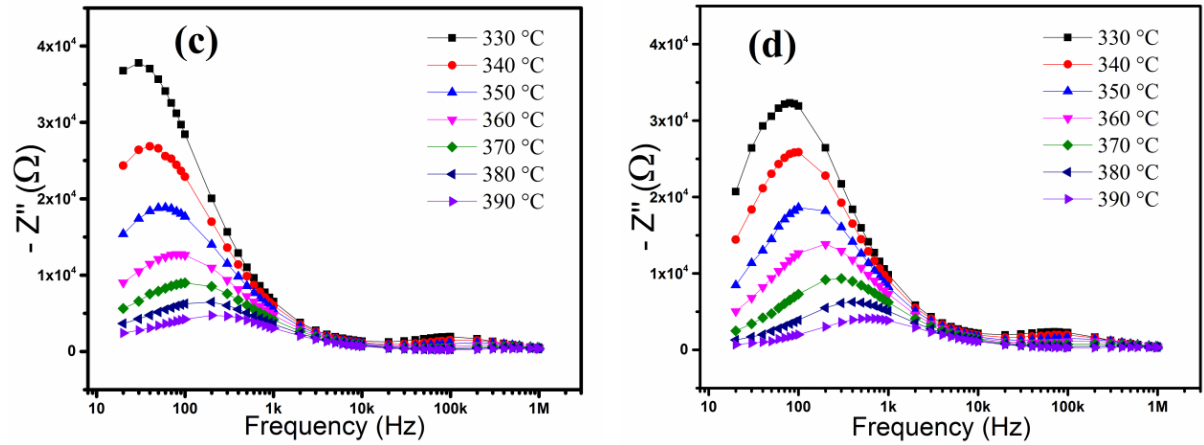

Fig.7. (Color online) Variation of real and imaginary part of complex impedance (Z*) with frequency at different temperatures for BMZT1 (a and c) and BMZT2 (b and d) samples.

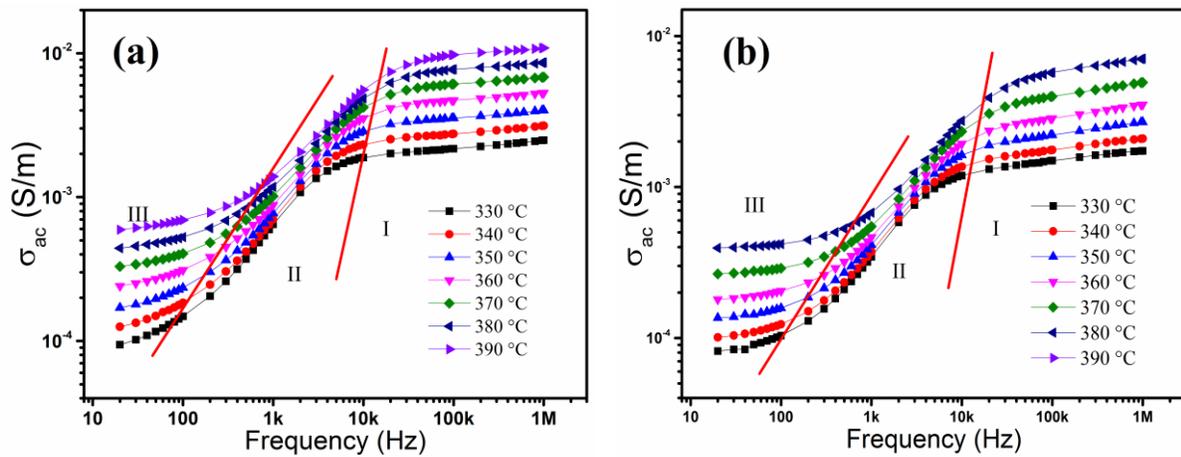

Fig.8. (Color online) Variation of ac conductivity as function of frequency at different temperatures of (a) BMZT1 and (b) BMZT2.

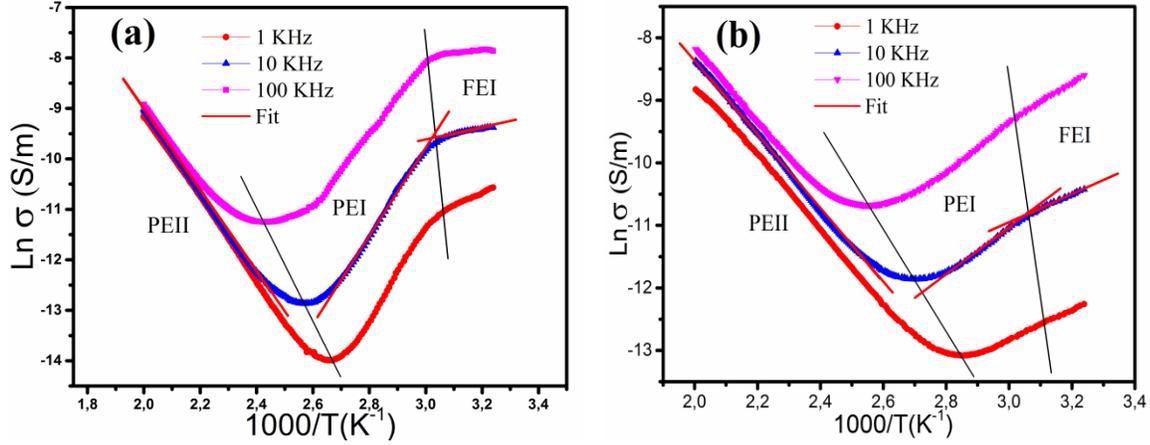

Figure.9. (Color online) Variation of ac conductivity vs inverse of temperature at different frequencies for (a) BMZT1 and (b) BMZT2.

## Tables

Table.1. Structural Parameters for (a) BMZT1 and (b) BMZT2 samples.

|   | x | y | z | Biso | Occ |
|---|---|---|---|---|---|
| (a) Lattice parameters: a=b= 4.0279 Å, c= 4.0234 Å, V= 65.2768 Å$^3$. Crystal system: Tetragonal. Space group: *P4mm*  $\chi^2$= 1.75, Rp=23.6, Rwp=23.1 | | | | | |
| Ba | 0.00000 | 0.00000 | 0.15928 | 0.74589 | 0.98685 |
| Mg | 0.00000 | 0.00000 | 0.15928 | 0.17011 | 0.01748 |
| Ti | 0.50000 | 0.50000 | 0.70452 | 0.54568 | 0.87905 |
| Zr | 0.50000 | 0.50000 | 0.70452 | 0.18832 | 0.09449 |
| $O_1$ | 0.50000 | 0.50000 | 0.22226 | 0.28338 | 2.06677 |
| $O_2$ | 0.00000 | 0.50000 | 0.62230 | 0.61957 | 1.36910 |
| (b) Lattice parameters: a=b= 4.0268Å, c= 4.0196Å, V= 65.1780Å$^3$. Crystal system: Tetragonal. Space group: *P4mm*  $\chi^2$= 1.47, Rp=21.4, Rwp=20.7 | | | | | |
| Ba | 0.00000 | 0.00000 | 0.00737 | 0.66214 | 0.97293 |
| Mg | 0.00000 | 0.00000 | 0.00737 | 0.25386 | 0.01983 |
| Ti | 0.50000 | 0.50000 | 0.44326 | 0.43148 | 0.89631 |
| Zr | 0.50000 | 0.50000 | 0.44326 | 0.30252 | 0.09439 |
| $O_1$ | 0.50000 | 0.50000 | 0.08253 | 0.00010 | 2.26638 |
| $O_2$ | 0.00000 | 0.50000 | 0.58245 | 0.23950 | 0.70594 |

Table.2. Santos-Eiras results of (a) BMZT1 and (b) BMZT2 samples.

| Frequency (a) | $\varepsilon_m$ | $T_m$ (°C) | Δ | γ |
|---|---|---|---|---|
| 100 HZ | 7904 | 56 | 31 | 1.75 |
| 1 kHz | 7410 | 57 | 31 | 1.76 |
| 100 kHz | 7066 | 58 | 32 | 1.76 |

| Frequency (b) | $\varepsilon_m$ | $T_m$ (°C) | Δ | γ |
|---|---|---|---|---|
| 100 HZ | 3819 | 38 | 58 | 1.95 |
| 1 kHz | 3674 | 39 | 55 | 1.99 |
| 100 kHz | 3540 | 40 | 56 | 1.97 |

Table.3. Resistance, capacitance values and relaxation time determined for bulk and grain boundary of (a) BMZT1 and (b) BMZT2 samples.

| T(°C) (a) | $R_g$ (Ω) | $Q_g$ (pF) | $n_1$ | $\tau_g$ (s*$10^{-6}$) | $R_{gb}$ (Ω) | $Q_{gb}$ (nF) | $n_2$ | $\tau_{gb}$ (s*$10^{-4}$) |
|---|---|---|---|---|---|---|---|---|
| 330 | 3679 | 444 | 0.98626 | 1.360 | 101830 | 130 | 0.79936 | 44.9 |
| 340 | 2867 | 435 | 0.98546 | 1.020 | 74321 | 151 | 0.78979 | 34.0 |
| 350 | 2195 | 429 | 0.98434 | 0.755 | 53480 | 176 | 0.78103 | 25.4 |
| 360 | 1632 | 425 | 0.98278 | 0.541 | 36822 | 206 | 0.77305 | 18.0 |
| 370 | 1247 | 423 | 0.98130 | 0.401 | 26477 | 238 | 0.76582 | 13.4 |
| 380 | 975.1 | 423 | 0.97975 | 0.305 | 19362 | 273 | 0.75992 | 10.1 |
| 390 | 762.4 | 425 | 0.97811 | 0.232 | 14170 | 318 | 0.75309 | 7.66 |

| T(°C) (b) | $R_g$ (Ω) | $Q_g$ (pF) | $n_1$ | $\tau_g$ (s*$10^{-6}$) | $R_{gb}$ (Ω) | $Q_{gb}$ (nF) | $n_2$ | $\tau_{gb}$ (s*$10^{-4}$) |
|---|---|---|---|---|---|---|---|---|
| 330 | 4311 | 576 | 0.97913 | 1.890 | 87673 | 72 | 0.81541 | 20 |
| 340 | 3648 | 607 | 0.97354 | 1.550 | 69788 | 78 | 0.81192 | 16.3 |
| 350 | 2878 | 668 | 0.96522 | 1.200 | 51129 | 84.9 | 0.80915 | 12.0 |
| 360 | 2202 | 724 | 0.95692 | 0.874 | 37889 | 94.6 | 0.8034 | 9.03 |
| 370 | 1537 | 855 | 0.94642 | 0.611 | 25420 | 104 | 0.79993 | 6.00 |
| 380 | 1029 | 984 | 0.93634 | 0.396 | 17011 | 117 | 0.79503 | 3.98 |
| 390 | 677.1 | 1210 | 0.92216 | 0.251 | 11300 | 134 | 0.78839 | 2.65 |

Table.4. Activation energy values Ea in eV for ac conduction, calculated assuming an Arrhenius behavior for each of the marked regions in Fig. 8 for three chosen frequencies for (a) BMZT1 and (b) BMZT2 samples.

| (a)  | 1kHz (eV) | 10kHz (eV) | 100kHz (eV) |
|------|-----------|------------|-------------|
| FEI  | 0,22      | 0,10       | 0,04        |
| PEI  | 0,81      | 0,74       | 0,62        |
| PEII | 0,70      | 0,69       | 0,65        |

| (b)  | 1kHz (eV) | 10kHz (eV) | 100kHz (eV) |
|------|-----------|------------|-------------|
| FEI  | 0,19      | 0,20       | 0,26        |
| PEI  | 0,20      | 0,31       | 0,33        |
| PEII | 0,50      | 0,51       | 0,49        |